\documentclass[conference]{IEEEtran}
\usepackage{graphicx}
\usepackage{amssymb}
\usepackage{bm}
\usepackage{amsmath}
\usepackage{pgfplots}
\usepackage{epstopdf}
\pgfplotsset{compat=newest}
\pgfplotsset{plot coordinates/math parser=false}
\newlength\figureheight
\newlength\figurewidth

\usepackage{color,soul}
\newcounter{MYtempeqncnt}
\newcommand\ccounter{18}
\IEEEoverridecommandlockouts
  
\begin{document}

\title{How Much Do Downlink Pilots Improve Cell-Free Massive MIMO?}
\author{\IEEEauthorblockN{Giovanni Interdonato$^{*\dagger}$, Hien Quoc Ngo$^\dagger$, Erik G. Larsson$^\dagger$, P{\aa}l Frenger$^*$}
\IEEEauthorblockA{$^*$Ericsson Research, Wireless Access Networks, 581 12 Link\"oping, Sweden\\
$^\dagger$Department of Electrical Engineering (ISY), Link\"oping University, 581 83 Link\"oping, Sweden\\
\{giovanni.interdonato, pal.frenger\}@ericsson.com, \{hien.ngo, erik.g.larsson\}@liu.se\\
\thanks{This paper was supported by the European Union's Horizon 2020 research
and innovation programme under grant agreement No 641985 (5Gwireless).}}}
\maketitle

\begin{abstract}
In this paper, we analyze the benefits of including downlink pilots in
a cell-free massive MIMO system. We derive an approximate per-user achievable
downlink rate for conjugate beamforming processing, which takes into
account both uplink and downlink channel estimation errors, and power
control. A performance comparison is carried out, in terms of per-user
net throughput, considering cell-free massive MIMO operation with and
without downlink training, for different network densities. We take
also into account the performance improvement provided by max-min
fairness power control in the downlink. Numerical results show that, exploiting downlink pilots, the performance can be considerably improved in low density networks over the conventional scheme where the users rely on statistical channel knowledge only. In high density networks, performance improvements are moderate. 
\end{abstract}


\section{Introduction}

Cell-Free massive multiple-input multiple-output (MIMO) refers to a
massive MIMO system \cite{MarzettaNonCooperative} where
the base station antennas are geographically distributed
\cite{NgoCF,MarzettaCF,Truong}.  These antennas, called access points (APs)
herein, simultaneously serve many users in the same frequency
band. The distinction between cell-free massive MIMO and conventional
distributed MIMO \cite{ZhouWCS} is the number of antennas involved in
coherently serving a given user.  In canonical cell-free massive MIMO,
every antenna serves every user. Compared to co-located massive MIMO,
cell-free massive MIMO has the potential to improve coverage and
energy efficiency, due to increased macro-diversity gain.


By operating in time-division duplex (TDD) mode, cell-free massive
MIMO exploits the channel reciprocity property, according to which the
channel responses are the same in both uplink and
downlink. Reciprocity calibration, to the required accuracy, can be
achieved in practice using off-the-shelf methods \cite{Lund}.  Channel
reciprocity allows the APs to acquire channel state information (CSI)
from pilot sequences transmitted by the users in the uplink, and this
CSI is then automatically valid also for the downlink.  By virtue of
the law of large numbers, the effective scalar channel gain seen by
each user is close to a deterministic constant. This is called
\textit{channel hardening}. Thanks to the channel hardening, the users
can reliably decode the downlink data using only statistical CSI.
This is the reason for why most previous studies on massive MIMO
assumed that the users do not acquire CSI and that there are no pilots
in the downlink
\cite{MarzettaNonCooperative,DebbahULDL,BjornsonHowMany}.
In co-located massive MIMO, transmission of downlink pilots and the
associated channel estimation by the users yields rather modest
performance improvements, owing to the high degree of channel
hardening \cite{NgoDlPilots,Khansefid,Zuo}.  In contrast, in cell-free
massive MIMO, the large number of APs is distributed over a wide area,
and many APs are very far from a given user; hence, each user is
effectively served by a smaller number of APs. As a result, the
channel hardening is less pronounced than in co-located massive MIMO,
and potentially the gain from using downlink pilots is larger.

\textbf{Contributions:} We propose a downlink training scheme for
cell-free massive MIMO, and provide an (approximate) achievable
downlink rate for conjugate beamforming processing, valid for finite
numbers of APs and users, which takes channel estimation errors and
power control into account. This rate expression facilitates a
performance comparison between cell-free massive MIMO with downlink
pilots, and cell-free massive MIMO without downlink pilots, where only
statistical CSI is exploited by the users.  The study is restricted to
the case of mutually orthogonal pilots, leaving the general case with
pilot reuse for future work.

\textit{Notation:} Column vectors are denoted by boldface letters. The superscripts $()^*$, $()^T$, and $()^H$ stand for the conjugate, transpose, and conjugate-transpose, respectively. The Euclidean norm and the expectation operators are denoted by $\Vert\cdot\Vert$ and $\mathbb{E}\{\cdot\}$, respectively. Finally, we use $z\sim\mathcal{CN}(0,\sigma^2)$ to denote a circularly symmetric complex Gaussian random variable (RV) $z$ with zero mean and variance $\sigma^2$, and use $z\sim\mathcal{N}(0,\sigma^2)$ to denote a real-valued Gaussian RV.


\section{System Model and Notation}

Let us consider $M$ single-antenna APs\footnote{We are considering the conjugate beamforming scheme which is  implemented in a distributed manner, and hence, an $N$-antenna APs can be treated as $N$ single-antenna APs.}, randomly spread out in a large
area without boundaries, which simultaneously serve $K$ single-antenna
users, $M>K$, by operating in TDD mode. All APs cooperate via a
backhaul network exchanging information with a central processing unit
(CPU). Only payload data and power control coefficients are
exchanged. Each AP locally acquires CSI and precodes data signals
without sharing CSI with the other APs. 
The time-frequency resources are divided into coherence intervals of length $\tau$ symbols (which are equal to the coherence time times the coherence bandwidth). The channel is assumed to be static within a coherence interval, and it varies independently between every coherence interval.

Let $g_{mk}$ denote the channel coefficient between the $k$th user
and the $m$th AP, defined as
\begin{equation}
\label{eq:channelmodel}
g_{mk} = \sqrt{\beta_{mk}}h_{mk},
\end{equation}
where $h_{mk}$ is the small-scale fading, and $\beta_{mk}$ represents
the large-scale fading. Since the APs are not co-located, the
large-scale fading coefficients $\{\beta_{mk}\}$ depend on both $m$
and $k$. We assume that $h_{mk}$, $m=1,..., M$, $k=1,..., K$, are i.i.d.\ $\mathcal{CN}(0,1)$ RVs,
i.e. Rayleigh fading. Furthermore, $\beta_{mk}$ is constant with
respect to frequency and is known, a-priori, whenever
required. Lastly, we consider moderate and low user mobility, thus
viewing $\{\beta_{mk}\}$ coefficients as constants.

The TDD coherence interval is divided into four phases: uplink
training, uplink payload data transmission, downlink training, and
downlink payload data transmission. In the uplink training phase,
users send pilot sequences to the APs and each AP estimates the
channels to all users. The channel estimates are used by the APs
to perform the uplink signal detection, and to beamform pilots and
data during the downlink training and the downlink data
transmission phase, respectively. Here, we focus on the the
downlink performance. The analysis on the uplink payload data
transmission phase is omitted, since it does not affect on the downlink performance.

\subsection{Uplink Training}

Let $\tau_\textrm{u,p}$ be the uplink training duration per
coherence interval such that $\tau_\textrm{u,p}<\tau$. Let
$\sqrt{\tau_\textrm{u,p}}\bm{\varphi}_k \in
\mathbb{C}^{\tau_\textrm{u,p}\times1}$, be the pilot sequence of
length $\tau_\textrm{u,p}$ samples sent by the $k$th user,
$k=1,...,K$. We assume that users transmit pilot sequences with full power, and all the uplink pilot sequences are mutually orthonormal, i.e., $\bm{\varphi}_k^H\bm{\varphi}_{k'} = 0$ for $k' \neq k$, and $\Vert\bm{\varphi}_k\Vert^2=1$. This
requires that $\tau_\textrm{u,p} \geq K$, i.e., $\tau_\textrm{u,p} = K$ is the smallest number of samples required to generate $K$ orthogonal vectors.

The $m$th AP receives a $ \tau_\textrm{u,p}\times1 $ vector of $K$ uplink pilots
linearly combined as
\begin{equation}
\label{eq:receiveduplinkpilot}
\textbf{y}_{\textrm{up},m} = \sqrt{\tau_\textrm{u,p}\rho_\textrm{u,p}}\sum^K_{k=1} g_{mk}\bm{\varphi}_k + \textbf{w}_{\textrm{up},m},
\end{equation}
where $\rho_\textrm{u,p}$ is the normalized transmit signal-to-noise
ratio (SNR) related to the pilot symbol and
$\textbf{w}_{\textrm{up},m}$ is the additive noise vector, whose elements are i.i.d. $\mathcal{CN}(0,1)$ RVs.

The $m$th AP processes the received pilot signal as follows
\begin{equation}
\label{eq:receiveduplinkpilotprojection}
\check{y}_{\textrm{up},mk} = \bm{\varphi}^H_k\textbf{y}_{\textrm{up},m} = \sqrt{\tau_\textrm{u,p}\rho_\textrm{u,p}} \ g_{mk}+ \bm{\varphi}^H_k\textbf{w}_{\textrm{up},m},
\end{equation}
and estimates the channel $g_{mk}$, $k=1,...,K$ by performing MMSE
estimation of $g_{mk}$ given $\check{y}_{\textrm{up},mk}$, which is
given by
\begin{equation}
\label{eq:APchannelestimation}
\hat{g}_{mk} = \frac{\mathbb{E}\{\check{y}^*_{\textrm{up},mk}g_{mk}\}}{\mathbb{E}\{\vert\check{y}_{\textrm{up},mk}\vert^2\}}\check{y}_{\textrm{up},mk} = c_{mk}\check{y}_{\textrm{up},mk},
\end{equation}
where
\begin{equation}
\label{eq:cmk}
c_{mk} \triangleq \frac{\sqrt{\tau_\textrm{u,p}\rho_\textrm{u,p}}\beta_{mk}}{\tau_\textrm{u,p}\rho_\textrm{u,p}\beta_{mk}+1}.
\end{equation}
The corresponding channel estimation error is denoted by
$\tilde{g}_{mk} \triangleq g_{mk} - \hat{g}_{mk}$ which is
independent of $\hat{g}_{mk}$.

\subsection{Downlink Payload Data Transmission}

During the downlink data transmission phase, the APs exploit the
estimated CSI to precode the signals to be transmitted to the $K$
users. Assuming conjugate beamforming, the transmitted
signal from the $m$th AP is given by
\begin{equation}
\label{eq:APtransmittedsignal}
x_m = \sqrt{\rho_\textrm{d}}\sum^K_{k=1} \sqrt{\eta_{mk}} \ \hat{g}^*_{mk} q_k,
\end{equation}
where $q_k$ is the data symbol intended for the $k$th user, which
satisfies $\mathbb{E}\{\vert q_k \vert^2\}=1$, and $\rho_\textrm{d}$
is the normalized transmit SNR related to the data symbol. Lastly,
$\eta_{mk}$, $m=1,...,M$, $k=1,...,K$, are power control coefficients
chosen to satisfy the following average power constraint at each AP:
\begin{equation}
\label{eq:pwConstraint}
\mathbb{E}\{|x_m|^2\}\leq\rho_\textrm{d}.
\end{equation}
Substituting (\ref{eq:APtransmittedsignal}) into (\ref{eq:pwConstraint}), the power constraint above can be rewritten as
\begin{equation}
\label{eq:pwConstraintGamma}
\sum \limits_{k=1}^K \eta_{mk} \gamma_{mk} \leq 1, \ \text{for all}\ m,
\end{equation}
where
\begin{equation}
\label{eq:defGamma}
\gamma_{mk} \triangleq \mathbb{E}\{{|\hat{g}_{mk}|}^2\} = \sqrt{\tau_\textrm{u,p}\rho_\textrm{u,p}} \beta_{mk} c_{mk}
\end{equation}
represents the variance of the channel estimate. The $k$th user
receives a linear combination of the data signals transmitted by
all the APs. It is given by
\begin{align}
\label{eq:receiveddownlinksignal}
r_{\textrm{d},k} &= \sum^M_{m=1} g_{mk} x_m + w_{\textrm{d},k} = \sqrt{\rho_\textrm{d}} \sum^K_{k'=1} a_{kk'} q_{k'} + w_{\textrm{d},k},
\end{align}
where
\begin{align}\label{eq:akkdef}
a_{kk'} \triangleq \sum^M_{m=1} \sqrt{\eta_{mk'}} {g}_{mk}
\hat{g}^*_{mk'}, ~ k'=1,...,K,
\end{align}
 and $ w_{\textrm{d},k} $ is
additive $\mathcal{CN}(0,1)$ noise at the $k$th user. In order to
reliably detect the data symbol $q_k$, the $k$th user must have a
sufficient knowledge of the effective channel gain, $a_{kk}$.

\begin{figure*}[!t]
\normalsize
\setcounter{MYtempeqncnt}{\value{equation}}

\setcounter{equation}{\ccounter}
\begin{equation}
\label{eq:genericDLrate}
R_k = \mathbb{E}\left\{\log_2\left(1+\frac{\rho_\textrm{d} \left| \mathbb{E}\left\{a_{kk} \mathrel{\big|} \hat{a}_{kk} \right\} \right|^2}{\rho_\textrm{d} \sum\limits^K_{k'=1} \mathbb{E}\left\{{|a_{kk'}|^2 \mathrel{\big|} \hat{a}_{kk}}\right\}-\rho_\textrm{d} \left| \mathbb{E}\left\{a_{kk} \mathrel{\big|} \hat{a}_{kk} \right\} \right|^2 +1}  \right)\right\}.
\end{equation}
\setcounter{equation}{\value{MYtempeqncnt}}
\hrulefill
\vspace*{4pt}
\end{figure*}

\subsection{Downlink Training}

While the model given so far is identical to that in \cite{NgoCF}, we
now depart from that by the introduction of downlink
pilots. Specifically, we adopt the Beamforming Training scheme proposed in \cite{NgoDlPilots}, where pilots are beamformed to the users. This scheme is scalable in that it does not require any information exchange among APs, and its channel estimation overhead is independent of $M$.

Let $\tau_\textrm{d,p}$ be the length (in symbols) of the downlink
training duration per coherence interval such that
$\tau_\textrm{d,p}<\tau - \tau_\textrm{u,p}$. The $m$th AP
precodes the pilot sequences $\bm{\psi}_{k'} \in
\mathbb{C}^{\tau_\textrm{d,p}\times1}$, $k'=1,...,K$, by using
the channel estimates $\{\hat{g}_{mk'}\}$, and beamforms it to all the
users. The $\tau_\textrm{d,p} \times 1$ pilot vector
$\bm{x}_{m,\textrm{p}}$ transmitted from the $m$th AP is given by
\begin{equation}
\label{eq:DLpilot} \bm{x}_{m,\textrm{p}} =
\sqrt{\tau_\textrm{d,p}\rho_\textrm{d,p}}\sum^K_{k'=1}
\sqrt{\eta_{mk'}} \hat{g}^*_{mk'} \bm\psi_{k'},
\end{equation}
where $\rho_\textrm{d,p}$ is the normalized transmit SNR per
downlink pilot symbol, and $\{\bm\psi_{k}\}$ are mutually
orthonormal, i.e. $\bm{\psi}^H_k\bm{\psi}_{k'} = 0$, for $k' \neq
k$, and $\Vert\bm{\psi}_k\Vert^2=1$. This requires that
$\tau_\textrm{d,p} \geq K$.

The $k$th user receives a corresponding $ \tau_\textrm{d,p}\times
1 $ pilot vector:
\begin{equation}
\label{eq:receivedDLpilot}
\textbf{y}_{\textrm{dp},k} = \sqrt{\tau_\textrm{d,p}\rho_\textrm{d,p}} \sum^K_{k'=1} a_{kk'} \bm\psi_{k'} + \textbf{w}_{\textrm{dp},k},
\end{equation}
where $\textbf{w}_{\textrm{dp},k}$ is a vector of additive noise
at the $k$th user, whose elements are i.i.d. $\mathcal{CN}(0,1)$
RVs.

In order to estimate the effective channel gain $a_{kk}$, $k=1,...,K$,
the $k$th user first processes the received pilot as
\begin{align}
\label{eq:receivedDLpilotprojection}
\check{y}_{\textrm{dp},k} &= \bm{\psi}^H_{k} \textbf{y}_{\textrm{dp},k} = \sqrt{\tau_\textrm{d,p}\rho_\textrm{d,p}} \ a_{kk} + \bm\psi^H_{k} \textbf{w}_{\textrm{dp},k} \nonumber \\
&= \sqrt{\tau_\textrm{d,p} \rho_\textrm{d,p}} \ a_{kk} + n_{\textrm{p},k},
\end{align}
where $n_{\textrm{p},k} \triangleq
\bm{\psi}^H_k\textbf{w}_{\textrm{dp},k} \sim
\mathcal{CN}(0,1)$, and then performs linear MMSE estimation of
$a_{kk}$ given $\check{y}_{\textrm{dp},k}$, which is, according to
\cite{SMKay}, equal to
\begin{align}
\label{eq:a_kk}
\hat{a}_{kk} &= \mathbb{E}\{a_{kk}\} + \nonumber \\
&+ \frac{\sqrt{\tau_\textrm{d,p}\rho_\textrm{d,p}} \ \mathrm{Var}\{a_{kk}\}}{\tau_\textrm{d,p} \rho_\textrm{d,p} \mathrm{Var}\{a_{kk}\} + 1}(\check{y}_{\textrm{dp},k} - \sqrt{\tau_\textrm{d,p}\rho_\textrm{d,p}} \ \mathbb{E}\{a_{kk}\}).
\end{align}
\textit{Proposition 1:} With conjugate beamforming, the linear MMSE
estimate of the effective channel gain formed by the $k$th user, see
(\ref{eq:a_kk}), is
\begin{align}
\label{eq:a_kk2}
\hat{a}_{kk} = \frac{\sqrt{\tau_\textrm{d,p}\rho_\textrm{d,p}}\ \varsigma_{kk} \ \check{y}_{\textrm{dp},k} + \sum \limits_{m=1}^M \sqrt{\eta_{mk}} \  \gamma_{mk}}{\tau_\textrm{d,p}\rho_\textrm{d,p}\ \varsigma_{kk} + 1},
\end{align}
where $\varsigma_{kk} \triangleq \sum_{m=1}^M \eta_{mk} \beta_{mk} \gamma_{mk}$.
\begin{IEEEproof}
See Appendix A.
\end{IEEEproof}

%

\begin{figure*}[!t]
\normalsize
\setcounter{MYtempeqncnt}{\value{equation}}
\setcounter{equation}{23}
\begin{align}
\label{eq:rateComparison}
R_k =
\begin{cases}
\log_2 \left( 1 + \frac{\rho_\textrm{d}\left(\sum\limits^M_{m=1} \sqrt{\eta_{mk}}\gamma_{mk}\right)^2}{\rho_\textrm{d} \sum\limits^K_{k'=1} \sum\limits^M_{m=1} \eta_{mk'} \beta_{mk} \gamma_{mk'} + 1}\right)&\text{for statistical CSI,} \\
(\ref{eq:DLrateApprox2})&\text{for Beamforming Training,}\\
\mathbb{E} \left\{ \log_2 \left( 1 + \frac{\rho_\textrm{d}|a_{kk}|^2}{\rho_\textrm{d} \sum\limits^K_{k' \neq k} |a_{kk'}|^2 + 1} \right)\right\}&\text{for perfect CSI.}
\end{cases}
\end{align}
\setcounter{equation}{\value{MYtempeqncnt}}
\hrulefill
\vspace*{4pt}
\end{figure*}

\section{Achievable Downlink Rate}

In this section we derive an achievable downlink rate for conjugate
beamforming precoding, using downlink pilots via Beamforming Training. An achievable downlink rate for the $k$th user is
obtained by evaluating the mutual information between the observed
signal $r_{\textrm{d},k}$ given by (\ref{eq:receiveddownlinksignal}),
the known channel estimate $\hat{a}_{kk}$ given by (\ref{eq:a_kk2})
and the unknown transmitted signal $q_k$:
$I(q_k;r_{\textrm{d},k},\hat{a}_{kk})$, for a permissible choice of
input signal distribution.

\begin{figure}[!t]
\centering
\includegraphics[width=3.3in]{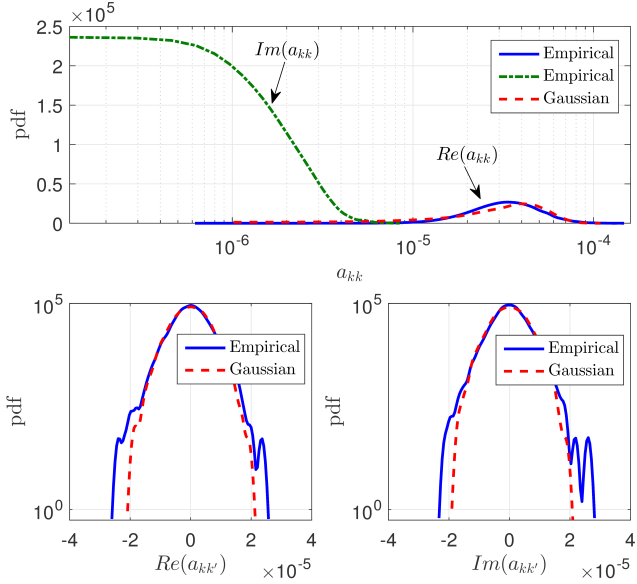}
\caption{The approximate (Gaussian) and the true (empirical) pdfs of
  $a_{kk}$ and $a_{kk'}$ for a given $\beta_{mk}$ realization (the
  large-scale fading model is discussed in detail in Section
  \ref{NumericalResults}). Here, $M = 20$ and $K = 5$.}
\label{pdfplot}
\end{figure}

Letting $\tilde{a}_{kk}$ be the channel estimation error, the effective
channel gain $a_{kk}$ can be decomposed as
\begin{equation}
\label{eq:ChEstErr}
a_{kk} = \hat{a}_{kk}+\tilde{a}_{kk}.
\end{equation}
Note that, since we use the linear MMSE estimation, the estimate
$\hat{a}_{kk}$ and the estimation error $\tilde{a}_{kk}$ are
uncorrelated, but not independent. The received signal at the
$k$th user described in (\ref{eq:receiveddownlinksignal}) can be
rewritten as
\begin{align}
r_{\textrm{d},k} = \sqrt{\rho_\textrm{d}}\ {a}_{kk} q_{k} +
\tilde{w}_{\textrm{d},k},
\end{align}
where $ \tilde{w}_{\textrm{d},k} \triangleq
\sqrt{\rho_\textrm{d}}\ \sum^K_{k' \neq k} a_{kk'} q_{k'} +
w_{\textrm{d},k} $ is the effective noise, which  satisfies
$\mathbb{E}\left\{\tilde{w}_{\textrm{d},k} \mathrel{\big|}
\hat{a}_{kk}
\right\}=\mathbb{E}\left\{q_k^\ast\tilde{w}_{\textrm{d},k}
\mathrel{\big|} \hat{a}_{kk} \right\}=\mathbb{E}\left\{a_{kk}^\ast
q_k^\ast\tilde{w}_{\textrm{d},k} \mathrel{\big|} \hat{a}_{kk}
\right\} = 0$. Therefore, following a similar methodology as in
\cite{Medard}, we obtain an achievable downlink rate of the
transmission from the APs to the $k$th user, which is given by
(\ref{eq:genericDLrate}) at the top of the page. The
expression given in (\ref{eq:genericDLrate}) can be simplified by
making the approximation that $a_{kk'}$,
$k'=1,...,K$, are Gaussian RVs. Indeed, according to the
Cram\'{e}r central limit theorem\footnote{Cram\'{e}r central limit
theorem: Let $X_1, X_2, ..., X_n$ are independent circularly
symmetric complex RVs. Assume that $X_i$ has zero mean and
variance $\sigma^2_i$. If $s^2_n = \sum^n_{i=1} \sigma^2_i
\rightarrow \infty$ and $\sigma_i/s_n \rightarrow 0$, as
$n\rightarrow \infty$, then $\frac{\sum^n_{i=1} X_i}{s_n}
\xrightarrow{d} \mathcal{CN}(0,1), \ \text{as } n \rightarrow
\infty$.}, we have \setcounter{equation}{19}
\begin{align}
\label{eq:approxAk1}
&a_{kk'} = \sum^M_{m=1} \sqrt{\eta_{mk'}} \ {g}_{mk} \hat{g}^*_{mk'} \xrightarrow{d} \mathcal{CN}\left(0, \varsigma_{kk'} \right), \text{ as } M \rightarrow \infty, \\
\label{eq:approxAkk}
&a_{kk}
    =
    \sum^M_{m=1} \sqrt{\eta_{mk}}  |\hat{g}_{mk}|^2  + \sum^M_{m=1} \sqrt{\eta_{mk}} \tilde{g}_{mk} \hat{g}^*_{mk} \nonumber \\
    &\approx
    \sum^M_{m=1} \sqrt{\eta_{mk}}  |\hat{g}_{mk}|^2 \xrightarrow{d} \mathcal{N}\left(\sum^M_{m=1} \sqrt{\eta_{mk}} \gamma_{mk},\sum^M_{m=1} \eta_{mk}  \gamma_{mk}^2 \right), \nonumber \\
    &\text{as } M \rightarrow \infty,
\end{align}
where $\varsigma_{kk'} \triangleq \sum_{m=1}^M \eta_{mk'} \beta_{mk}
\gamma_{mk'}$, and $\xrightarrow{d}$ denotes convergence in
distribution. The Gaussian approximations (\ref{eq:approxAk1}) and
(\ref{eq:approxAkk}) can be verified by numerical results, as shown in
Figure \ref{pdfplot}. The pdfs show a close match between the
empirical and the Gaussian distribution even for small
$M$. Furthermore, with high probability the imaginary part of $a_{kk}$
is much smaller than the real part so it can be reasonably neglected.

Under the assumption that $a_{kk}$ is Gaussian distributed,
$\hat{a}_{kk}$ in (\ref{eq:a_kk2}) becomes the MMSE estimate of
$a_{kk}$. As a consequence, $\hat{a}_{kk}$ and $\tilde{a}_{kk}$ are
independent. In addition, by following a similar methodology as in
\eqref{eq:approxAk1} and \eqref{eq:approxAkk}, we can show that any
linear combination of $a_{kk}$ and $a_{kk'}$ are asymptotically (for
large $M$) Gaussian distributed, and hence $a_{kk}$ and $a_{kk'}$ are
asymptotically jointly Gaussian distributed. Furthermore, $a_{kk}$ and
$a_{kk'}$ are uncorrelated so they are independent. Hence, the achievable downlink rate (\ref{eq:genericDLrate}) is reduced to\footnote{ A formula similar to \eqref{eq:DLrateApprox} but for co-located massive MIMO systems, was given in \cite{NgoDlPilots,Khansefid} with equality between the left and right hand sides.  Those expressions were not rigorously correct capacity lower bounds (although very good approximations), as $a_{kk}$ is non-Gaussian in general. }
\begin{equation}
\label{eq:DLrateApprox} R_k \!\approx\! \mathbb{E}\! \left\{\!
\log_2 \!\left( \!1 \!+\!
\frac{\rho_\textrm{d}|\hat{a}_{kk}|^2}{\rho_\textrm{d}
\mathbb{E}\{|\tilde{a}_{kk}|^2\} + \rho_\textrm{d}
\sum\limits^K_{k' \neq k} \mathbb{E}\{|a_{kk'}|^2\} + 1} \!\right)
\!\right\}.
\end{equation}

\textit{Proposition 2:} With conjugate beamforming, an achievable
rate of the transmission from the APs to the $k$th user is
\begin{align}
\label{eq:DLrateApprox2}
R_k \approx \mathbb{E} \left\{ \log_2 \left( 1 + \frac{\rho_\textrm{d}|\hat{a}_{kk}|^2}{\rho_\textrm{d}\frac{\varsigma_{kk}}{\tau_\textrm{d,p}\rho_\textrm{d,p}\varsigma_{kk}+1} + \rho_\textrm{d} \sum\limits^K_{k' \neq k} \varsigma_{kk'} + 1} \right) \right\}.
\end{align}

\begin{IEEEproof}
See Appendix B.
\end{IEEEproof}

\section{Numerical Results}
\label{NumericalResults}

We compare the performance of cell-free massive MIMO for three
different assumptions on CSI: \textit{(i)} Statistical CSI, without
downlink pilots and users exploiting only statistical knowledge of the
channel gain \cite{NgoCF}; \textit{(ii)} Beamforming Training,
transmitting downlink pilots and users estimating the gain from those
pilots; \textit{(iii)} Perfect CSI, where the users know the effective
channel gain. The latter represents an upper bound (genie) on
performance, and is not realizable in practice.
The gross spectral efficiencies for these cases are given by (\ref{eq:rateComparison}) at the top of the page.

Taking into account the performance loss due to the downlink and uplink pilots, the \textit{per-user net throughput} (bit/s) is
\setcounter{equation}{24}
\begin{equation} 
\mathcal{S}_k = B \frac{1-\tau_{\textrm{oh}}/\tau}{2} R_k,
\end{equation}
where $B$ is the bandwidth, $\tau$ is the length of the coherence interval in samples, and $\tau_{\textrm{oh}}$ is the pilots overhead, i.e., the number of samples per coherence interval spent for the training phases.

We further examine the performance improvement by using the
\textit{max-min fairness power control} algorithm in \cite{NgoCF},
which provides equal and hence uniformly good service to all users for
the Statistical CSI case. When using this algorithm for the
Beamforming Training case (and for the Perfect CSI bound), we use the
power control coefficients computed for the Statistical CSI case. This
is, strictly speaking, not optimal but was done for computational
reasons, as the rate expressions with user CSI are not in closed form.

\subsection{Simulation Scenario}

Consider $M$ APs and $K$ users uniformly randomly distributed within a
square of size $1 \text{ km}^2$. The large-scale fading coefficient $\beta_{mk}$ is modeled as
\begin{equation}
\label{eq:beta}
\beta_{mk} = \text{PL}_{mk} \cdot 10^{\frac{\sigma_{sh}z_{mk}}{10}}
\end{equation}  
where $\text{PL}_{mk}$ represents the path loss, and $10^{\frac{\sigma_{sh}z_{mk}}{10}}$ is the shadowing with standard deviation $\sigma_{sh}$ and $z_{mk}\sim\mathcal{N}(0,1)$.
We consider the three-slope model for the path loss as in \cite{NgoCF} and uncorrelated shadowing. We adopt the following parameters:
the carrier frequency is 1.9 GHz, the bandwidth is 20 MHz, the shadowing standard deviation is 8 dB, and the noise figure (uplink and downlink) is 9 dB. In all examples (except for Figures~\ref{cdfM50K10hpw} and \ref{cdfM50K10dpw}) the radiated power (data and pilot) is 200 mW for APs and 100 mW for users. The corresponding normalized transmit SNRs can be computed by dividing radiated powers by the noise power, which is given by
\begin{align*}
\text{noise power} = \text{bandwidth} \times k_B \times T_0 \times \text{noise figure} \text{ (W)},
\end{align*}
where $k_B$ is the Boltzmann constant, and $T_0 = 290$ (Kelvin) is the noise temperature.
The AP and user antenna height is 15 m, 1.65 m, respectively. The antenna gains are $0$ dBi. Lastly, we take $\tau_{\textrm{d,p}} = \tau_{\textrm{u,p}} = K$, and $\tau = 200$
samples which corresponds to a coherence bandwidth of 200 kHz and a
coherence time of 1 ms. To avoid cell-edge effects, and to imitate a
network with an infinite area, we performed a wrap-around technique,
in which the simulation area is wrapped around such that the nominal
area has eight neighbors.

\subsection{Performance Evaluation}

We focus first on the performance gain, over the conventional scheme,
provided by jointly using Beamforming Training scheme and max-min
fairness power control in the downlink. We consider two scenarios, with
different network densities.
\begin{figure}[!t]
\centering
\includegraphics[width=3.3in]{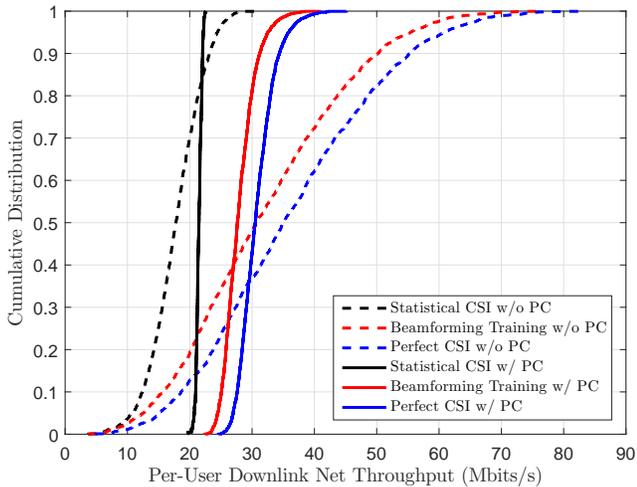}
\caption{The cumulative distribution of the per-user downlink net throughput with and without max-min power control (PC), for the case of statistical, imperfect and perfect CSI knowledge at the user, $M = 50$ and $K = 10$.}
\label{cdfM50K10}
\end{figure}
Figure \ref{cdfM50K10} shows the cumulative distribution function (cdf)
of the per-user net throughput for the three cases, with $M=50$,
$K=10$. In such a low density scenario, the channel hardening is less pronounced and
performing the Beamforming Training scheme yields high performance
gain over the statistical CSI case. Moreover, the Beamforming Training
curve approaches the upper bound. Combining max-min power control with
Beamforming Training scheme, gains can be further improved. For
instance, Beamforming Training provides a performance improvement of
$18\%$ over the statistical CSI case in terms of 95\%-likely per-user
net throughput, and $29\%$ in terms of median per-user net
throughput. 

\begin{figure}[!t]
\centering
\includegraphics[width=3.3in]{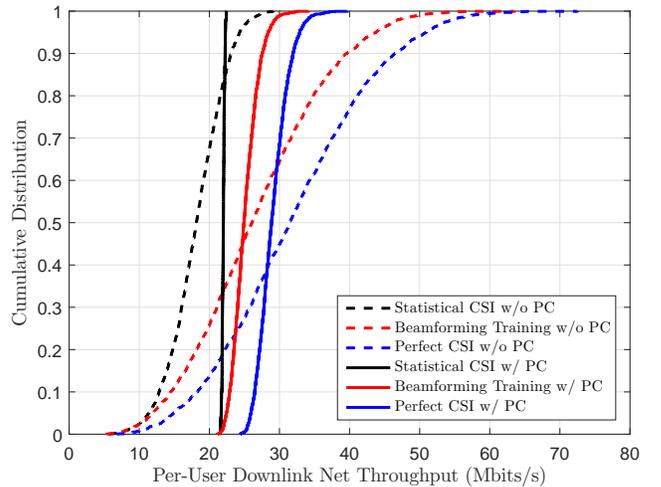}
\caption{The cumulative distribution of the per-user downlink net throughput with and without max-min power control (PC), for the case of statistical, imperfect and perfect CSI knowledge at the user, $M = 100$ and $K = 20$.}
\label{cdfM100K20}
\end{figure}

By contrast, for higher network densities the gap between
statistical and Beamforming Training tends to be reduced due to two
factors: $(i)$ as $M$ increases, the statistical CSI knowledge at the
user side is good enough for reliable downlink detection due to the
channel hardening; $(ii)$ as $K$ increases, the pilot overhead becomes
significant. In Figure \ref{cdfM100K20} the scenario with $M=100$,
$K=20$ is illustrated. Here, the 95\%-likely and the median per-user
net throughput of the Beamforming Training improves of $4\%$ and
$13\%$, respectively, the performance of the statistical CSI case.

Max-min fairness power control maximizes the rate of the worst
user. This philosophy leads to two noticeable consequences: $(i)$ the
curves describing with power control are more concentrated around
their medians; $(ii)$ as $K$ increases, performing power control has
less impact on the system performance, since the probability to have users experiencing poor channel conditions increases.

\begin{figure}[!t]
\centering
\includegraphics[width=3.3in]{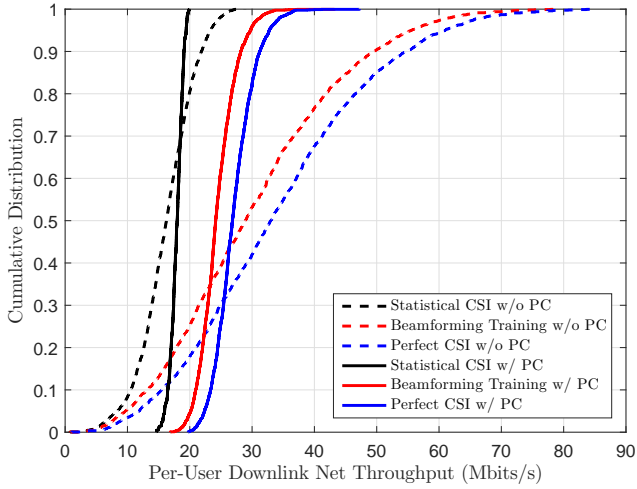}
\caption{The same as Figure \ref{cdfM50K10}, but the radiated power for data and pilot is 50 mW for APs and 20 mW for users.}
\label{cdfM50K10hpw}
\end{figure}

\begin{figure}[!t]
\centering
\includegraphics[width=3.3in]{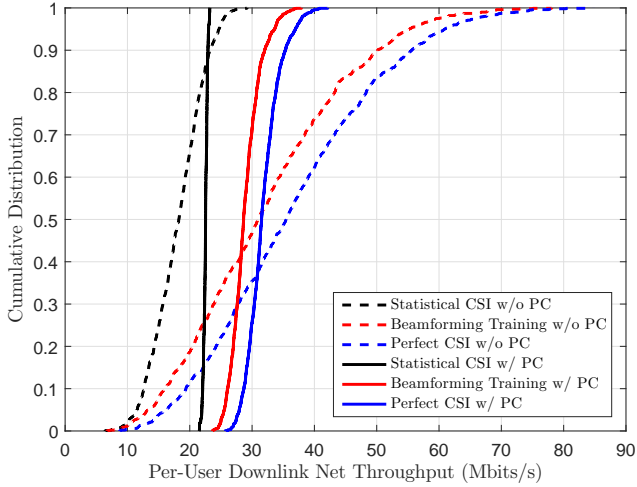}
\caption{The same as Figure \ref{cdfM50K10}, but the radiated power for data and pilot is 400 mW for APs and 200 mW for users.}
\label{cdfM50K10dpw}
\end{figure}

Finally, we compare the performance provided by the two schemes by setting different values for the radiated powers. In Figure \ref{cdfM50K10hpw}, the radiated power is set to 50 mW and 20 mW for the downlink and the uplink, respectively, with $M=50$ and $K=10$. In low SNR regime, with max-min fairness power control, Beamforming Training scheme outperforms the statistical CSI case of about $26\%$ in terms of 95\%-likely per-user net throughput, and about $34\%$ in terms of median per-user net throughput.
Similar performance gaps are obtained by increasing the radiated power to 400 mW for the downlink and 200 mW for the uplink, as shown in Figure \ref{cdfM50K10dpw}.  


\section{Conclusion}

Co-located massive MIMO systems do not need downlink training
since by virtue of channel hardening, the effective channel gain
seen by each user fluctuates only slightly around its mean. In
contrast, in cell-free massive MIMO, only a small number of APs
may substantially contribute, in terms of transmitted power, to
serving a given user, resulting in less channel hardening.  We
showed that by transmitting downlink pilots, and performing
Beamforming Training together with max-min fairness power control,
performance of cell-free massive MIMO can be substantially
improved. 

We restricted our study to the case of mutually
orthogonal pilots. The general case with non-orthogonal pilots may
be included in future work. Further work may also include pilot assignment algorithms, optimal power control, and the analysis of zero-forcing precoding technique.

\section*{Appendix}
\subsection{Proof of Proposition 1}
\begin{itemize}
\item Compute $\mathbb{E}\{a_{kk'}\}$:

From \eqref{eq:akkdef}, and by using $g_{mk}\triangleq\hat{g}_{mk}+\tilde{g}_{mk}$, we have
\begin{align}
\label{eq:ChEstWithErr}
& a_{kk'} = \sum^M_{m=1} \sqrt{\eta_{mk'}} \hat{g}_{mk} \hat{g}^*_{mk'} + \sum^M_{m=1} \sqrt{\eta_{mk'}} \tilde{g}_{mk} \hat{g}^*_{mk'}.
\end{align}
Owing to the properties of MMSE estimation, $\tilde{g}_{mk}$ and
$\hat{g}_{mk}$ are independent, $k=1, \ldots, K$. Therefore,
\begin{align}
\label{eq:Eakk}
\mathbb{E}\{a_{kk'}\} &= \mathbb{E}\left\{\sum^M_{m=1} \sqrt{\eta_{mk'}} \hat{g}_{mk} \hat{g}^*_{mk'}\right\} \nonumber \\
&=
  \begin{cases}
    0       & \quad \text{if } k' \neq k\\
    \sum\limits^M_{m=1} \sqrt{\eta_{mk}} \ \gamma_{mk}  & \quad \text{if } k' = k.\\
  \end{cases}
\end{align}

\item Compute $\mathrm{Var}\{a_{kk}\}$:
\begin{equation}
\label{eq:Varakk}
\mathrm{Var}\{a_{kk}\} = \mathbb{E}\{{|a_{kk}|}^2\} - {|\mathbb{E}\{a_{kk}\}|}^2.
\end{equation}
According to (\ref{eq:ChEstWithErr}), we get
\begin{align}
\label{eq:EakkSq}
& \mathbb{E}\{{|a_{kk}|}^2\} = \mathbb{E} \left\{\left| \sum\limits^M_{m=1} \sqrt{\eta_{mk}} |\hat{g}_{mk}|^2 \right|^2\right\}  \nonumber \\
&\qquad\qquad\qquad\qquad {} + \mathbb{E} \left\{\left| \sum\limits^M_{m=1} \sqrt{\eta_{mk}} \tilde{g}_{mk} \hat{g}^*_{mk} \right|^2 \right\}   \nonumber \\
&\stackrel{(a)}{=} \mathbb{E} \left\{ \sum\limits^M_{m=1} \sum\limits^M_{m'=1} \sqrt{\eta_{mk}} |\hat{g}_{mk}|^2 \sqrt{\eta_{m'k}} |\hat{g}_{m'k}|^2 \right\}  \nonumber \\
&\qquad\qquad {} + \sum\limits^M_{m=1} \eta_{mk}(\beta_{mk} - \gamma_{mk})\gamma_{mk} \nonumber \\
&= \sum\limits^M_{m=1} \sum\limits^M_{m'=1} \sqrt{\eta_{mk} \eta_{m'k}} \  \mathbb{E} \left\{ |\hat{g}_{mk}|^2 |\hat{g}_{m'k}|^2 \right\} + \nonumber \\
&\qquad\qquad {} + \sum\limits^M_{m=1} \eta_{mk}(\beta_{mk} - \gamma_{mk})\gamma_{mk} \nonumber \\
&= \sum\limits^M_{m=1} \eta_{mk}(\beta_{mk} - \gamma_{mk})\gamma_{mk} + \sum\limits^M_{m=1} \eta_{mk} \ \mathbb{E} \left\{ |\hat{g}_{mk}|^4 \right\}  \nonumber \\
&\qquad\qquad {} + \sum\limits^M_{m=1} \sum\limits^M_{m' \neq m} \sqrt{\eta_{mk} \eta_{m'k}} \  \mathbb{E} \left\{ |\hat{g}_{mk}|^2 |\hat{g}_{m'k}|^2 \right\} \nonumber \\
&\stackrel{(b)}{=} \sum\limits^M_{m=1} \eta_{mk}(\beta_{mk} - \gamma_{mk})\gamma_{mk} + 2 \sum\limits^M_{m=1} \eta_{mk} \gamma_{mk}^2  \nonumber \\
&\qquad\qquad {} + \sum\limits^M_{m=1} \sum\limits^M_{m' \neq m} \sqrt{\eta_{mk} \eta_{m'k}} \  \gamma_{mk} \gamma_{m'k},
\end{align}
where $(a)$ follows from the fact that $\mathbb{E}\{{|\tilde{g}_{mk}|}^2\} = \beta_{mk} - \gamma_{mk}$, and $(b)$ from $\mathbb{E}\left\{{\left|\hat{g}_{mk} \right|}^4\right\} = 2 \gamma^2_{mk}$.

From (\ref{eq:Eakk}), we have
\begin{align}
\label{eq:sqEakk}
& {|\mathbb{E}\{a_{kk}\}|}^2 = \left| \sum\limits^M_{m=1} \sqrt{\eta_{mk}} \gamma_{mk} \right|^2 \nonumber \\
&= \sum\limits^M_{m=1} \sum\limits^M_{m'=1} \sqrt{\eta_{mk} \eta_{m'k}} \ \gamma_{mk} \gamma_{m'k} \nonumber \\
&= \sum\limits^M_{m=1} \eta_{mk} \gamma_{mk}^2 + \sum\limits^M_{m=1} \sum\limits^M_{m' \neq m} \sqrt{\eta_{mk} \eta_{m'k}} \  \gamma_{mk} \gamma_{m'k}  .
\end{align}
Substituting (\ref{eq:EakkSq}) and (\ref{eq:sqEakk}) into (\ref{eq:Varakk}), we obtain
\begin{equation}
\label{eq:VarakkVal}
\mathrm{Var}\{a_{kk}\} = \sum\limits^M_{m=1} \eta_{mk} \beta_{mk} \gamma_{mk} = \varsigma_{kk}.
\end{equation}
Substituting (\ref{eq:Eakk}) and (\ref{eq:VarakkVal}) into (\ref{eq:a_kk}), we get (\ref{eq:a_kk2}).
\end{itemize}

\subsection{Proof of Proposition 2}

\begin{itemize}
\item Compute $\mathbb{E}\{{|a_{kk'}|}^2\}$ for $k' \neq k$:

From (\ref{eq:ChEstWithErr}) and (\ref{eq:Eakk}), we have
\begin{align}
\label{eq:Varaki}
&\mathbb{E}\{{|a_{kk'}|}^2\} = \mathrm{Var}\{a_{kk'}\} \nonumber \\
&= \mathbb{E}\left\{{\left|\sum\limits^M_{m=1} \sqrt{\eta_{mk'}} \hat{g}_{mk} \hat{g}^*_{mk'}\right|}^2\right\}  \nonumber \\
&\qquad\qquad\qquad\qquad {} + \mathbb{E}\left\{{\left|\sum\limits^M_{m=1} \sqrt{\eta_{mk'}} \tilde{g}_{mk} \hat{g}^*_{mk'}\right|}^2\right\} \nonumber \\
&= \sum\limits^M_{m=1} \eta_{mk'} \mathbb{E}\left\{{\left|\hat{g}_{mk} \hat{g}^*_{mk'}\right|}^2\right\} + \sum\limits^M_{m=1} \eta_{mk'} \mathbb{E}\left\{{\left|\tilde{g}_{mk} \hat{g}^*_{mk'}\right|}^2\right\} \nonumber \\
&\stackrel{(a)}{=} \sum\limits^M_{m=1} \eta_{mk'} \gamma_{mk} \gamma_{mk'} + \sum\limits^M_{m=1} \eta_{mk'}(\beta_{mk} - \gamma_{mk})\gamma_{mk'} \nonumber \\
&= \sum\limits^M_{m=1} \eta_{mk'} \beta_{mk} \gamma_{mk'} = \varsigma_{kk'},
\end{align}
where $(a)$ is obtained by using (\ref{eq:defGamma}) and the fact
that $\tilde{g}_{mk}$ has zero mean and is independent of
$\hat{g}_{mk}$. Moreover, we have that
$\mathbb{E}\{{|\tilde{g}_{mk}|}^2\} = \beta_{mk} - \gamma_{mk}$.

\item Compute $\mathbb{E}\{{|\tilde{a}_{kk}|}^2\}$:

From (\ref{eq:a_kk2}) and (\ref{eq:ChEstErr}), we have
\begin{align}
\label{eq:ekk2}
& \mathbb{E}\{{|\tilde{a}_{kk}|}^2\} = \mathbb{E}\{{|a_{kk}-\hat{a}_{kk}|}^2\} \nonumber \\
&= \mathbb{E}\left\{\left|a_{kk} - \frac{\sqrt{\tau_\textrm{d,p}\rho_\textrm{d,p}} \varsigma_{kk} \check{y}_{\textrm{dp},k}}{\tau_\textrm{d,p}\rho_\textrm{d,p}\varsigma_{kk} + 1} - \frac{\sum_{m=1}^M \sqrt{\eta_{mk}}  \gamma_{mk}}{\tau_\textrm{d,p}\rho_\textrm{d,p} \varsigma_{kk} + 1} \right|^2\right\} \nonumber \\
&\stackrel{(a)}{=} \mathbb{E}\left\{\left|a_{kk} \left(1- \frac{\tau_\textrm{d,p}\rho_\textrm{d,p} \varsigma_{kk}}{\tau_\textrm{d,p}\rho_\textrm{d,p}\varsigma_{kk} + 1}\right) - \frac{\sum_{m=1}^M \sqrt{\eta_{mk}}  \gamma_{mk}}{\tau_\textrm{d,p}\rho_\textrm{d,p} \varsigma_{kk} + 1}  \right. \right. \nonumber \\
&\quad\quad  \left. \left. - \frac{\sqrt{\tau_\textrm{d,p}\rho_\textrm{d,p}} \varsigma_{kk} n_{\textrm{p},k}}{\tau_\textrm{d,p}\rho_\textrm{d,p}\varsigma_{kk} + 1} \right|^2\right\} \nonumber \\
&= \mathbb{E}\left\{\left|\frac{a_{kk} - \sum_{m=1}^M \sqrt{\eta_{mk}}  \gamma_{mk} - \sqrt{\tau_\textrm{d,p}\rho_\textrm{d,p}} \varsigma_{kk} n_{\textrm{p},k}}{\tau_\textrm{d,p}\rho_\textrm{d,p}\varsigma_{kk} + 1} \right|^2\right\} \nonumber \\
&\stackrel{(b)}{=} \frac{\mathbb{E}\left\{\left|a_{kk} - \mathbb{E}\{a_{kk}\} - \sqrt{\tau_\textrm{d,p}\rho_\textrm{d,p}}\ \varsigma_{kk} n_{\textrm{p},k} \right|^2\right\}}{\left(\tau_\textrm{d,p}\rho_\textrm{d,p} \varsigma_{kk}+1\right)^2} \nonumber \\
&\stackrel{(c)}{=} \frac{\mathrm{Var}\{a_{kk}\} + \tau_\textrm{d,p}\rho_\textrm{d,p} \varsigma^2_{kk}}{\left(\tau_\textrm{d,p}\rho_\textrm{d,p} \varsigma_{kk}+1\right)^2} \nonumber \\
&= \frac{\varsigma_{kk} + \tau_\textrm{d,p}\rho_\textrm{d,p} \varsigma^2_{kk}}{\left(\tau_\textrm{d,p}\rho_\textrm{d,p} \varsigma_{kk}+1\right)^2}   \nonumber \\
&= \frac{\varsigma_{kk}}{\tau_\textrm{d,p}\rho_\textrm{d,p} \varsigma_{kk}+1},
\end{align}
where $(a)$ is obtained by using (\ref{eq:receivedDLpilotprojection}), and $(b)$ by using (\ref{eq:Eakk}). Instead, $(c)$ follows from the fact that $a_{kk} - \mathbb{E}\left\{a_{kk}\right\}$, $n_{\textrm{p},k}$ are independent and zero-mean RVs. Moreover, $n_{\textrm{p},k}$ has unitary variance.

Substituting (\ref{eq:Varaki}) and (\ref{eq:ekk2}) in (\ref{eq:DLrateApprox}), we arrive at the result in Proposition 2.

\end{itemize}
\bibliographystyle{IEEEtran}
\bibliography{IEEEabrv,mybib}
\end{document}